\documentclass[twocolumn,twocolappendix]{aastex631}
\usepackage{CJK}

\usepackage[T1]{fontenc}

\usepackage{multirow}
\usepackage{xcolor}

\begin{document}
\begin{CJK*}{UTF8}{ipxm}

\title{Asymmetric dust accumulation of the PDS 70 disk revealed by ALMA Band 3 observations}

\author[0000-0003-1958-6673]{Kiyoaki Doi (土井聖明)}
\affiliation{Department of Astronomical Science, School of Physical Sciences, The Graduate University for Advanced Studies, SOKENDAI, 2-21-1 Osawa, Mitaka, Tokyo 181-8588, Japan}
\affiliation{National Astronomical Observatory of Japan, 2-21-1 Osawa, Mitaka, Tokyo 181-8588, Japan}
\correspondingauthor{Kiyoaki Doi}
\email{doi.kiyoaki.astro@gmail.com}

\author[0000-0003-4562-4119]{Akimasa Kataoka (片岡章雅)}
\affiliation{Department of Astronomical Science, School of Physical Sciences, The Graduate University for Advanced Studies, SOKENDAI, 2-21-1 Osawa, Mitaka, Tokyo 181-8588, Japan}
\affiliation{National Astronomical Observatory of Japan, 2-21-1 Osawa, Mitaka, Tokyo 181-8588, Japan}

\author[0000-0003-2300-2626]{Hauyu Baobab Liu (呂浩宇)}
\affiliation{Department of Physics, National Sun Yat-Sen University, No. 70, Lien-Hai Road, Kaohsiung City 80424, Taiwan, R.O.C.}
\affiliation{Center of Astronomy and Gravitation, National Taiwan Normal University, Taipei 116, Taiwan}

\author[0000-0001-8002-8473]{Tomohiro C. Yoshida (吉田有宏)}
\affiliation{Department of Astronomical Science, School of Physical Sciences, The Graduate University for Advanced Studies, SOKENDAI, 2-21-1 Osawa, Mitaka, Tokyo 181-8588, Japan}
\affiliation{National Astronomical Observatory of Japan, 2-21-1 Osawa, Mitaka, Tokyo 181-8588, Japan}

\author[0000-0002-7695-7605]{Myriam Benisty}
\affiliation{Max-Planck Institute for Astronomy, Königstuhl 17, D-69117 Heidelberg, Germany}
\affiliation{Universit\'{e} C\^{o}te d'Azur, Observatoire de la C\^{o}te d'Azur, CNRS, Laboratoire Lagrange, F-06304 Nice, France}
\affiliation{Universit\'{e} Grenoble Alpes, CNRS, IPAG, 38000 Grenoble, France}

\author[0000-0001-9290-7846]{Ruobing Dong (董若冰)}
\affiliation{Department of Physics \& Astronomy, University of Victoria, Victoria, BC, V8P 5C2, Canada}

\author[0000-0003-4099-6941]{Yoshihide Yamato (大和義英)}
\affiliation{Department of Astronomy, Graduate School of Science, The University of Tokyo, 7-3-1 Hongo, Bunkyo-ku, Tokyo 113-0033, Japan}

\author[0000-0002-3053-3575]{Jun Hashimoto (橋本淳)}
\affiliation{National Astronomical Observatory of Japan, 2-21-1 Osawa, Mitaka, Tokyo 181-8588, Japan}
\affiliation{Astrobiology Center, National Institutes of Natural Sciences, 2-21-1 Osawa, Mitaka, Tokyo 181-8588, Japan}

\begin{abstract}

The PDS 70 system, hosting two planets within its disk, is an ideal target for examining the effect of planets on dust accumulation, growth, and ongoing planet formation.
Here, we present high-resolution ($0\farcs07 = 8 \ \mathrm{au}$) dust continuum observations of the PDS 70 disk in ALMA Band 3 (3.0 mm).
While previous Band 7 observations showed a dust ring with slight asymmetry, our Band 3 observations reveal a more prominent asymmetric peak in the northwest direction, where the intensity is 2.5 times higher than in other directions and the spectral index is at the local minimum with $\alpha_{\mathrm{SED}} \sim 2.2$.
This indicates that a substantial amount of dust is accumulated both radially and azimuthally in the peak.
We also detect point-source emission around the stellar position in the Band 3 image, which is likely to be free-free emission.
We constrain the eccentricity of the outer ring to be $e<0.04$ from the position of the central star and the outer ring.
From the comparison with numerical simulations, we constrain the mass of PDS 70c to be less than 4.9 Jupiter masses if the gas turbulence strength $\alpha_{\mathrm{turb}} = 10^{-3}$.
Then, we discuss the formation mechanism of the disk structures and further planet formation scenarios in the disk.

\end{abstract}

\keywords{Protoplanetary disks (1300); Planet formation (1241); Submillimeter astronomy (1647); Dust continuum emission (412)}

\section{Introduction} \label{sec:intro}

Recent high-resolution millimeter observations with ALMA have revealed substructures of dust emission in many disks, such as rings \citep[e.g.,][]{ALMA2015,Andrews2018} and non-axisymmetric structures \citep[e.g.,][]{vanderMarel2013,Casassus2013,Fukagawa2013}, which may play an important role in dust growth and planet formation.
Local dust trapping may halt the radial drift of dust grains \citep[e.g.,][]{Whipple1972}.
Furthermore, the local enhancement of dust density may promote the growth of dust grains by collisional sticking \citep[e.g.,][]{Brauer2008} or lead to the formation of planetesimals through streaming instability \citep[e.g.,][]{Youdin2005,Johansen2007} or gravitational instability \citep[e.g.,][]{Sekiya1983,Youdin2002}.
Despite the importance of these structures, their formation mechanism is still unclear.

One of the possible mechanisms for the formation of these structures is the influence of planets in the disk.
Planets open gaps in the disk gas, and dust is trapped at the gas pressure maximum outside the planet, forming a dust ring \citep[e.g.,][]{Zhu2012,Pinilla2012a}.
If the gas gap is deep enough, vortices may form by the Rossby Wave Instability, and dust grains accumulate at the center of the vortex, leading to the formation of non-axisymmetric structures \citep[e.g.,][]{Lovelace1999,Ono2016}.
In addition, circumplanetary disks form around planets. 
These planets grow by mass accretion through the circumplanetary disks, and satellites are formed within these disks \citep[e.g.,][]{Ward2010,Tanigawa2012}.

PDS 70 is an ideal target for examining the interaction between the disk and the planets.
This object is a T Tauri star with an age of $5.4 \pm 1.0 \ \mathrm{Myr}$ \citep{Muller2018} at a distance of $112.39 \pm 0.24 \ \mathrm{pc}$ \citep{Gaia2023}.
Observations with the VLT/SPHERE and VLT/MUSE have discovered two accreting planets, PDS 70b and PDS 70c, at orbital radii of 22 and 34 au from the central star \citep{Keppler2018,Haffert2019,Hashimoto2020}.
ALMA Band 7 (0.87 mm) observations have detected continuum emission around PDS 70c and PDS 70b, possibly indicating circumplanetary disks \citep{Isella2019,Benisty2021,Casassus2022}.
Also, a wide dust ring at 74 au outside the planets' orbit and a faint inner disk near the central star have been detected \citep{Keppler2019}.
The cavity has also been detected in infrared observations \citep{Hashimoto2012,Dong2012} and molecular line observations \citep{Keppler2019,Law2024,Rampinelli2024}.

In this study, we characterize the dust distribution of PDS 70 using high-resolution ALMA Band 3 (3.0 mm) continuum observations. 
Previous high-resolution ALMA observations of PDS 70 have been limited to ALMA Band 7 (0.87 mm), which may be optically thick, making it difficult to characterize the disk dust properties.
At these wavelengths, the disks are often optically thick \citep{Tripathi2017,Ansdell2018,Chung2024}, and observations at longer wavelengths may reveal different structures from submillimeter observations \citep{Casassus2019, Liu2024}.
In this study, we present high-resolution observations at Band 3, which is likely more optically thin compared to Band 7, to reveal the detailed dust distribution of PDS 70 and characterize the dust properties.

The structure of this paper is as follows.
First, in Section \ref{sec:PDS70_obs}, we describe the observations and data analysis.
In Section \ref{sec:PDS70_results}, we present the observational images and the spectral index.
In Section \ref{sec:PDS70_discussion}, we discuss the central source, the circumplanetary disks, the mass of the planet in the disk, the dust mass in the ring, and the planet formation scenario.
Finally, we summarize the results in Section \ref{sec:PDS70_summary}.

\section{Observations} \label{sec:PDS70_obs}

The ALMA Band 3 observations of PDS 70 were conducted in Cycle 10 as two projects, 2022.1.00893.S (PI: K. Doi) and 2022.1.01477.S (PI: H. Liu).
The observations were performed in four configurations, C5, C6, C8 and C9, on March 26, May 4, July 2 and July 28, 2023, with on-source integration times of 2600, 756, 5334 and 3193 seconds, respectively.
The maximum recoverable scale of the C5 observations is $9\farcs83$, which is sufficient to avoid missing flux.
Four spectral windows were set at the central frequency of 90.5, 92.5, 102.5, and 104.5 GHz with a bandwidth of 1.875 GHz each to maximize the continuum sensitivity.
The phase calibrator was J1407-4302, and the amplitude calibrator was J1427-4206 for all observations.

The pipeline calibration was performed with Common Astronomical Software Applications (CASA) version 6.4.1.12 and the subsequent reduction was performed with CASA modular version 6.6.3.22 \citep{CASA2022}.
We subtracted the line emission and then performed phase-only self-calibration for each EB on the line-subtracted data.
The data were then combined and imaged using the \texttt{tclean} task. 
Imaging was performed with Multi-Frequency Synthesis (MFS) at a central frequency of 97.5 GHz, using \texttt{nterms = 2} and Briggs weighting with a robust parameter of 0.5. 
We did not apply self-calibration to the combined data because it did not improve SNR due to the insufficient peak SNR of 21.
The resulting image has a beam size of $0\farcs073 \times 0\farcs066$ with a position angle of $-64\fdg2$, and the rms noise level measured in the emission free region is $4.94\ \mathrm{\mu Jy\ beam^{-1}}$.

The Band 7 data are those shown in the left panel of Fig. 1 in \citet{Benisty2021}.
These data were obtained by 2015.1.00888.S (PI: E. Akiyama) and 2018.A.00030.S (PI: M. Benisty).
The data reduction is described in \citet{Benisty2021}.
We re-ran \texttt{tclean} on the calibrated data of \citet{Benisty2021} without applying the JvM correction \citep{JvM1995}, which was applied in \citet{Benisty2021}.
The resulting image has a beam size of $0\farcs046 \times 0\farcs035$ at a position angle of $-53\fdg1$, and the rms noise level is $12.7\ \mathrm{\mu Jy\ beam^{-1}}$ at the central frequency of 350.6 GHz.

\section{Results} \label{sec:PDS70_results}

\begin{figure*}[t]
  \centering
  \includegraphics[width=0.99\textwidth]{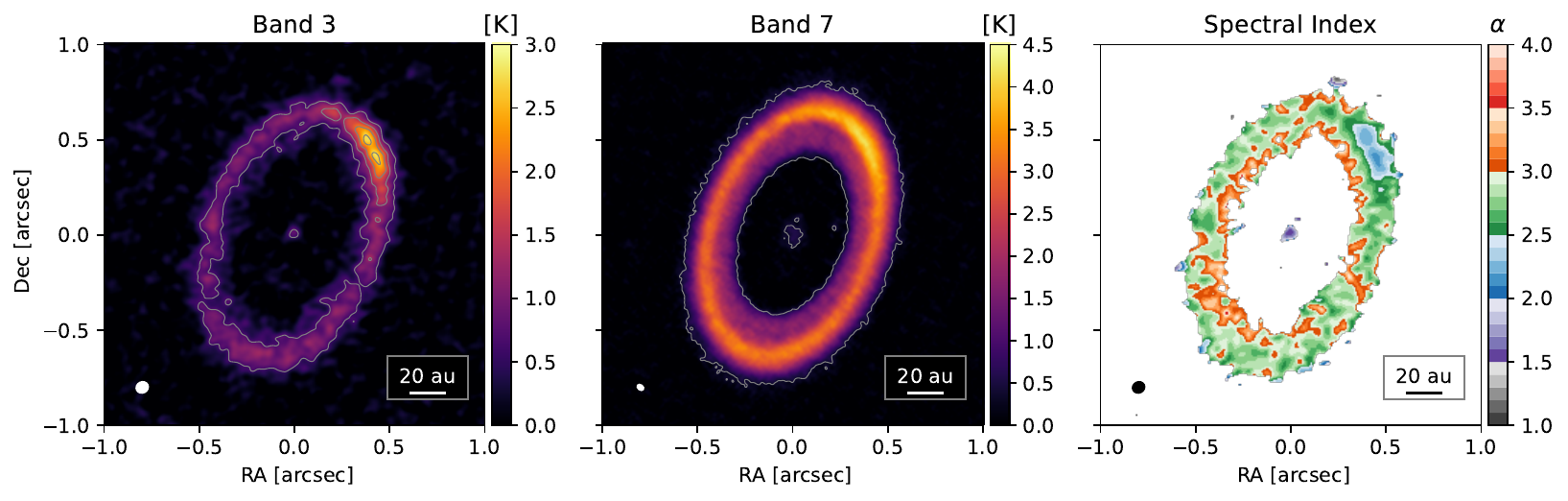}
    \caption{
      Images of PDS 70. Left: Band 3, Middle: Band 7, Right: Spectral energy distribution (SED) between Band 7 and 3.
      The intensity is shown in brightness temperature using the Rayleigh-Jeans approximation.
      The peak brightness temperature in Band 7 of 4.2 K using the Rayleigh-Jeans approximation corresponds to 10.4 K in the full Planck brightness temperature.
      The contours in the left panel show 5, 10, 15, and 20 times the rms noise level, and the contours in the middle panel show 5 times the rms noise level.
    }
  \label{fig:PDS70_images}
\end{figure*}

The left and middle panels of Figure \ref{fig:PDS70_images} show the images of PDS 70 in Band 3 and Band 7, respectively.
The Band 7 image shows three major structures: (1) a wide ring, (2) faint central emission, and (3) circumplanetary disks \citep{Keppler2019,Isella2019,Benisty2021,Casassus2022}.
In contrast, the Band 3 image shows a different morphology from that of Band 7.
First, the ring exhibits more prominent asymmetric peak in the northwest direction (hereafter, the northwest peak).
Second, the central emission shows a compact point-source emission, while it has a more extended structure in Band 7.
Third, the circumplanetary disk is not detected in Band 3.
We discuss the characteristics of these structures individually in Section \ref{sec:PDS70_discussion}.
To quantitatively estimate the distributions of these structures at each wavelength, we fit the ring and the central emission individually in Appendix \ref{app:ring_fit}.

The right panel of Figure \ref{fig:PDS70_images} shows the spectral index between Band 7 and Band 3.
The two images are aligned to match the center of the outer ring, which we derive by the fitting described in Appendix \ref{app:ring_fit}.
The Band 7 image is smoothed to match the resolution of the Band 3 beam.
The rms noise level of the smoothed Band 7 image is $15.1 \ \mathrm{\mu Jy\ beam^{-1}}$.
The spectral index is calculated as
\begin{equation}
  \alpha_{\mathrm{SED}} = \frac{\log(I_{\nu_{\mathrm{B7}}}/I_{\nu_{\mathrm{B3}}})}{\log(\nu_{\mathrm{B7}}/\nu_{\mathrm{B3}})}.
\end{equation}
where $I_{\nu_{\mathrm{B7}}}$ and $I_{\nu_{\mathrm{B3}}}$ are the intensity at the frequency $\nu_{\mathrm{B7}}$ and $\nu_{\mathrm{B3}}$, respectively.
The spectral index uncertainty between Band 7 and Band 3 resulting from absolute flux calibration uncertainties is $\Delta \alpha_{\mathrm{SED}} = 0.083$ when the absolute flux calibration error is 5\% for Band 3 and 10\% for Band 7 (ALMA Technical Handbook; \citealt{ALMA_TH2023}).
We note that flux calibration uncertainty affects the spectral index uniformly over the image but does not impact the difference in the spectral index between different regions. 
Therefore, unless otherwise noted, the spectral index uncertainty mentioned below only considers the uncertainty due to the rms noise, as we focus on the spatial variation of the spectral index.

The observed spectral index in the outer ring becomes a minimum at the northwest peak with $\alpha_{\mathrm{SED}} = 2.17 \pm 0.04$ with the rms error, and $\alpha_{\mathrm{SED}} = 2.17 \pm 0.09$ with flux calibration error.
The spectral index of the central point source is $\alpha_{\mathrm{SED}} = 1.53 \pm 0.13$ with the rms error, and $\alpha_{\mathrm{SED}} = 1.53 \pm 0.15$ with flux calibration error.

\section{Discussion} \label{sec:PDS70_discussion}

We characterize the dust ring in Section \ref{sec:PDS70_ring}, the central emission in Section \ref{sec:PDS70_central_source}, and the circumplanetary disk in Section \ref{sec:PDS70_circumplanetary_disk}.
We estimate the planet mass from the eccentricity of the ring in Section \ref{sec:PDS70_planet_mass}.
Finally, we discuss the formation process of the disk structure and planets in the PDS 70 system in Section \ref{sec:PDS70_planet_formation}.

\subsection{Ring} \label{sec:PDS70_ring}

The ring shows a more prominent northwest peak in Band 3 than in Band 7.
The spectral index takes a minimum value of $\alpha_{\mathrm{SED}} = 2.17 \pm 0.09$ at the northwest peak.
Since the spectral index approaches $\alpha_{\mathrm{SED}} \sim 2$ when the dust is optically thick, the observed low spectral index at the northwest peak suggests that the dust emission is optically thick, at least marginally.

\begin{figure}[hbtp]
  \centering
  \includegraphics[width=0.45\textwidth]{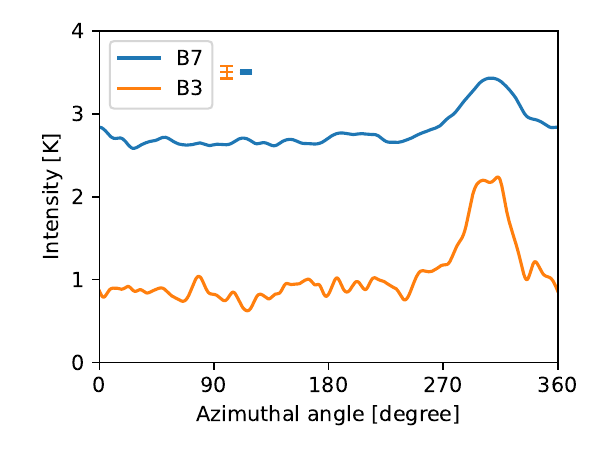}
    \caption{
      The azimuthal intensity profile along the ring in Band 7 (blue) and Band 3 (orange).
      The images are smoothed to be a circular beam with a size of $0\farcs12$ after deprojection.
      The error bar in the upper left indicates the $1 \sigma$ rms noise level.
    }
  \label{fig:PDS70_azimuthal}
\end{figure}

We show the azimuthal intensity profile along the ring in Band 7 and Band 3 in Figure \ref{fig:PDS70_azimuthal} to examine the azimuthal intensity variation.
To avoid the effect of the elliptical beam shape on the azimuthal variation of the intensity \citep{Doi2021}, we further smoothed both images so that the beam became a circular Gaussian after deprojection with a beam size of $0\farcs12$.
The rms noise level after the smoothing is $19.0 \ \mathrm{\mu Jy\ beam^{-1}}$ in Band 7 and $5.27 \ \mathrm{\mu Jy\ beam^{-1}}$ in Band 3.
The azimuthal variation of the intensity shows a significant difference between the two bands.
In Band 7, the intensity at the northwest peak is 1.3 times larger than in other directions, while it is 2.5 times larger in Band 3.
The smaller azimuthal contrast in Band 7 than in Band 3 implies that the dust is optically thick in all directions in Band 7, and the intensity is saturated with respect to the surface density.
On the other hand, the prominent azimuthal variation in Band 3 implies that the dust is optically thin in directions other than the northwest peak.
Thus, the Band 3 observations reveal that the dust is accumulated in the azimuthal direction at the northwest peak.

The Band 3 image tentatively shows a double peak structure at the northwest peak, but the intensity dip between the peaks is less than $3 \sigma$, making it difficult to conclusively characterize this feature. 
This feature may suggest dust circulation around a vortex \citep{Hammer2021} or the azimuthal accumulation of dust grains of different sizes \citep{Baruteau2016}.

\begin{figure}[htbp]
  \centering
  \includegraphics[width=0.45\textwidth]{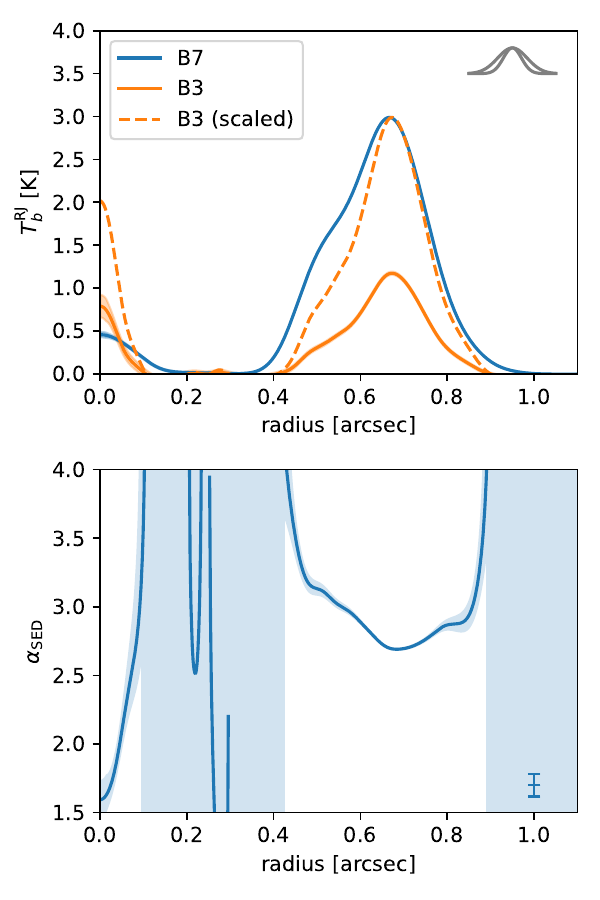}
    \caption{
      Top: The radial profile of the intensity in Band 7 (blue) and Band 3 (orange).
      The beam size is $0\farcs073 \times 0\farcs066$ for both wavelengths, which corresponds to $0\farcs113 \times 0\farcs069$ after deprojection.
      The dashed line represents the Band 3 profile scaled to match the peak intensity of Band 7 for comparison of the radial morphology.
      The shaded area represents the uncertainties, calculated as the rms noise level divided by the square root of the number of independent data points.
      The Gaussian in the upper right corner shows the major and minor axes of the beam after deprojection.
      Bottom: The radial profile of the spectral index.
      The error bar in the lower right corner indicates the $1\sigma$ error range due to the absolute flux calibration uncertainty. 
      The shaded region shows the uncertainties propagated from the uncertainties of the radial profiles.
    }
  \label{fig:PDS70_radial}
\end{figure}

The upper panel of Figure \ref{fig:PDS70_radial} shows the azimuthally averaged radial profile of the intensity, and the lower panel shows the radial profile of the spectral index.
To compare the profiles between the two bands, we smoothed the Band 7 image to have the same beam as Band 3 as in Section \ref{sec:PDS70_results}.
The orange dashed line in the upper panel is the Band 3 profile scaled to match the peak intensity of Band 7 for comparison of the radial morphology.
The radial profile is averaged over all azimuthal angles, including the northwest peak.
The radial profiles in the peak direction and the direction excluding the peak are shown in Appendix \ref{app:radial_profile}.
The error range is calculated as the rms noise level divided by the square root of the number of independent data points. 
Here, the number of independent data points is defined as the circumference divided by the geometric mean of the major and minor axes of the beam.

The upper panel shows that the dust ring width is narrower in Band 3 than in Band 7.
The fitting results in Appendix \ref{app:ring_fit} confirm that the ring width is narrower in Band 3 than in Band 7, with $w = 8.45_{-0.21}^{+0.22}$ au in Band 3 and $w = 11.23_{-0.03}^{+0.03}$ au in Band 7.
The lower panel shows that the spectral index has a minimum value of $\alpha_{\mathrm{SED}} = 2.69 \pm 0.01$ near the radial peak of the ring and increases toward the edge of the ring.
One explanation is that larger dust grains, which are more sensitive to long-wavelength observations, are more efficiently accumulated at the center of the ring, making the ring appear narrower at longer wavelengths.
The other explanation is that the peak intensity saturates due to the higher optical depth in Band 7 than in Band 3.

The shape of the ring is not symmetric between the inner and outer sides, and a structure called a shoulder is seen around $r \sim 0\farcs5$ in both wavelengths \citep{Keppler2019}.
Comparing the orange dashed line (scaled Band 3) and the blue line (Band 7) in Figure \ref{fig:PDS70_radial}, the intensity at the inner shoulder relative to the peak intensity is weaker in Band 3 than in Band 7.
Comparing the spectral index at $r = 0\farcs511$ and $r = 0\farcs770$, where the intensity is half of the peak intensity in Band 7, the spectral index is $\alpha_{\mathrm{SED}} = 3.12 \pm 0.05$ at the inner side and $\alpha_{\mathrm{SED}} = 2.81 \pm 0.03$ at the outer side.
This suggests that the dust size is smaller at the inner shoulder than the outer side.
\citet{Bi2024} show that the 3D gas flow induced by a planet forms a shoulder with small dust grains, but they predict the shoulder to be on the outer side, contrary to our observations. 
The inner shoulder structure can be explained if the dust grains leak inward from a loosely confined dust ring \citep{Jiang2021}.
Additionally, \citet{Pinilla2024} show that the shoulder may form if small dust grains leak from the ring over the gas gap.

We estimate the lower limit of dust mass in the ring from the Band 3 emission to discuss the possibility of further planet formation. 
We assumed a temperature of 16 K for the disk ring, based on the temperature model of \citet{Law2024}.
This is considered an upper limit model for the temperature, as it is derived from the upper layer of the gas, and the midplane temperature may be lower. 
We assumed the upper limit of the Band 3 dust opacity to be $\kappa_{\mathrm{abs}} = 0.47\ \mathrm{cm^2/g}$. 
Here, we used the DSHARP dust model for the dust opacity \citep{Birnstiel2018}, assuming a power-law size distribution of $q= 3.5$ \citep{Mathis1977,Doi2023} from $0.1 \ \mathrm{\mu m}$ grains to $3.4 \ \mathrm{mm}$ grains, in which the maximum dust size was chosen to maximize the opacity for conservative mass estimates.
We estimated the lower limit of the dust mass to be 28 $M_{\mathrm{Earth}}$ in the entire ring and 9.3 $M_{\mathrm{Earth}}$ even within the narrow range of $270^{\circ} < \phi < 360^{\circ}$ around the northwest peak.
Thus, the dust mass in the ring is sufficient to form a gas giant core of 10 $M_{\mathrm{Earth}}$ even with conservative estimates \citep{Mizuno1980,Ikoma2000}.

\subsection{Central Source} \label{sec:PDS70_central_source}

The central point source has a significantly different morphology depending on the wavelength.
As mentioned in Section \ref{sec:PDS70_results}, the spectral index of the central source has $\alpha_{\mathrm{SED}} < 2$, suggesting the contribution of the free-free emission \citep{Riviere-Marichalar2024,Rota2024}.
As shown in Figure \ref{fig:MCMC_B3_point}, the size of the central point source in Band 3 has an upper limit below the beam size, indicating compact emission.
On the other hand, the central emission in Band 7 is extended with a size of $\sim 8$ au in standard deviation as shown in Figure \ref{fig:MCMC_B7_point}.
Therefore, the compact central emission in Band 3 is dominated by free-free emission, while the extended emission in Band 7 suggests that it is dominated by dust thermal emission.
The presence of such free-free emission implies that accretion onto the central star is occurring \citep{Pascucci2012} onto the weak line T-Tauri star \citep{Gregorio-Hetem2002} with the relatively old age of 5.4 $\pm$ 1.0 Myr, the clear gap in both dust and gas \citep{Law2024}, and the disk wind \citep{Campbell-White2023}.

The flux estimated from the peak intensity of the central source in Band 3 is 31.1 $\mu$Jy.
\citet{Liu2024} analyzed multi-wavelength ALMA and JVLA data, and the spectral fit of the total flux showed that the emission from PDS 70 is dominated by dust emission at $> 30$ GHz and by  Anomalous Microwave Emission (AME) due to spinning dust at $< 30$ GHz, and free-free emission is not dominant at any frequency.
Although our study suggests that the central emission in Band 3 is dominated by free-free emission, it is much weaker than the total emission from the disk, which is consistent with the results of \citet{Liu2024}.

\subsection{Circumplanetary Disk} \label{sec:PDS70_circumplanetary_disk}

The previous Band 7 observations have detected a circumplanetary disk around PDS 70 c, and tentatively detected it around PDS 70 b \citep{Isella2019,Benisty2021,Casassus2022}.
Since the growth of gas giants occurs through circumplanetary disks, and satellite formation also occurs through circumplanetary disks, it is important to characterize the circumplanetary disks from their spectrum.

Our Band 3 observations did not detect any point source emission around PDS 70 b or c.
We put a lower limit on the spectral index of PDS 70 c as $\alpha_{\mathrm{SED}} > 1.5$, assuming the Band 7 intensity of PDS 70 c is $86 \pm 16$ $\mu$Jy \citep{Benisty2021} and the $3 \sigma$ upper limit in Band 3 is 15 $\mu$Jy.
The non-detection of the circumplanetary disk in Band 3 observation is not surprising, since the spectral index of thermal dust emission is $\alpha_{\mathrm{SED}} > 2$.

\subsection{Eccentricity and Exoplanet Mass} \label{sec:PDS70_planet_mass}

The planetary mass is a key parameter in understanding planet formation and disk structure.
The planet mass has been estimated from infrared observations, comparing the infrared photometry with planet evolution models.
The mass of PDS 70 c $M_{p,c}$ has been estimated to be $4 M_{\mathrm{Jupiter}} < M_{p,c} < 12 M_{\mathrm{Jupiter}}$ by \citet{Haffert2019}, $M_{p,c} < 5 M_{\mathrm{Jupiter}}$ by \citet{Mesa2019}, $1 M_{\mathrm{Jupiter}} < M_{p,c} < 3 M_{\mathrm{Jupiter}}$ by \citet{Wang2020}, and $M_{p,c} \sim 4 M_{\mathrm{Jupiter}}$ by \citet{Wahhaj2024}.
These estimates from infrared observations significantly depend on the assumptions of planet evolution models and the system age.
\citet{Portilla-Revelo2023} estimated the planet mass to be $M_{p,c} \sim 4 M_{\mathrm{Jupiter}}$ by comparing the gas gap depth from ALMA CO isotopic observations and the numerical simulation by \citet{Duffell2015}.
Overall, the mass of PDS 70 c remains uncertain, and constraints from different perspectives are needed.

The disk eccentricity, derived from millimeter continuum emission, offers a new way to constrain the planet mass.
More massive planets form deeper and wider gaps.
In particular, when the gap created by the planet is very deep, the ring-gap structure becomes unstable and gives eccentricity to the outer ring \citep{Kley2006,Tanaka2022}.
\citet{Bae2019} simulated a disk mimicking the PDS 70 system with a viscosity of $\alpha_{\mathrm{turb}} = 10^{-3}$ and introduced planets corresponding to PDS 70 b and c.
They showed that the ring becomes eccentric with the eccentricity $e = 0.2 - 0.4$ when the mass of PDS 70 c is 10 $M_{\mathrm{Jupiter}}$.
\citet{Tanaka2022} investigated the eccentricity of the outer ring induced by a single planet and derived empirical criteria for the critical planet mass $M_{\mathrm{crit}}$ for the ring to be eccentric with $e \gtrsim 0.07$, depending on the aspect ratio $h/r$, the turbulent strength $\alpha_{\mathrm{turb}}$, and the central star mass $M_{\mathrm{star}}$.
Using their equation (18)\footnote{
  \citet{Tanaka2022} present two equations (15) and (18) as critical planet masses, and state that it is currently unknown which one is better criteria from the current simulations. 
  Here, as an upper limit of the PDS 70 c, we used the equation (18), which is a conservative one when using the aspect ratio assumed for PDS 70.
}, the critical planet mass to induce the eccentricity is
\begin{equation}
  M_{\mathrm{crit}} =  4.9 M_{\mathrm{Jupiter}} \left( \frac{\alpha_{\mathrm{turb}}}{10^{-3}} \right)^{1/2}
\end{equation}
for the central star mass $M_{\mathrm{star}} = 0.85 M_{\odot}$ and the aspect ratio $h/r = 0.079$ \citep{Bae2019}.

The Band 3 observations allow us to accurately estimate the eccentricity by detecting point-source free-free emission at the position of the central star, i.e., the focus, as well as the outer ring. 
We can estimate the eccentricity from the offset between the position of the central star and the center of the ring, $\delta r$, and the semi-major axis $r_0$ as
\begin{equation}
  e = \frac{\delta r}{r_0}.
\end{equation}
As shown in Figure \ref{fig:MCMC_B3_center_shift} in Appendix \ref{app:ring_fit}, the center of the point source and the outer ring are consistent, indicating that the ring is circular with the eccentricity $e < 0.04$.
The offset between the center of the outer ring and the center of the central disk based on the Band 7 observations also shows that the centers are consistent, as shown in Figure \ref{fig:MCMC_B7_center_shift}, and provides the eccentricity of $e < 0.02$.
These low eccentricities put the mass of PDS 70 c below $4.9 M_{\mathrm{Jupiter}}$ for $\alpha_{\mathrm{turb}} = 10^{-3}$, which is proportional to the square root of the turbulent strength $\alpha_{\mathrm{turb}}$.
Thus, we put a strong upper limit on the planet mass from the new perspective of the disk eccentricity.
A caveat is that the simulation of \citet{Tanaka2022} is a 2D simulation assuming a single planet on a circular orbit, and the multiplicity or the eccentricity of the planets in PDS 70 or 3D effects may affect the eccentricity of the disk \citep{Wang2021,Li2023}.

\subsection{Planet Formation Scenario} \label{sec:PDS70_planet_formation}

We discuss the formation scenario of the disk structure and possible further planet formation in the PDS 70 system. 
The presence of the planets and the disk morphologies are consistent with the following scenario of planet-induced dust accumulation and planet formation.
First, the already-formed planets create the gap in the gas disk, trapping dust grains outside the gap \citep{Lau2024}.
The gas gap then induces vortices through the Rossby Wave Instability \citep{Lovelace1999,Ono2016}, possibly aided by disk winds \citep{Wu2023}.
The pressure bump in the gas accumulates dust radially, and the vortex accumulates dust azimuthally, leading to the formation of an asymmetric ring structure, as observed in Band 3.

The accumulated dust grains by the planets promote further planet formation.
As shown in the Band 3 image, the planets accumulate dust radially and azimuthally outside of the planets.
Locally accumulated dust grains can lead to the formation of planetesimals through the streaming instability \citep{Youdin2005,Johansen2007} or the gravitational instability \citep{Sekiya1983,Sekiya1998,Youdin2002}.
Planetesimals grow into solid planets and can form gas giants through gas accretion, and the Band 3 observations show that there is enough dust to form a gas giant core in PDS 70.
Thus, we suggest that inside-out planet formation may be occurring in PDS 70 \citep{Chatterjee2014,Tan2016}.
We note that these observations do not exclude other planet and structure formation scenarios \citep[e.g.,][]{Jiang2023}.

\section{Summary} \label{sec:PDS70_summary}

We performed high-resolution ALMA Band 3 (3 mm) observations of the PDS 70 system, a disk with planets.
We characterized the dust properties and the planets within the disk from the morphology and the spectral index of the Band 3 and the previous Band 7 (0.87 mm) images.
The main results are as follows:

\begin{enumerate}
  \item The dust ring has a clear northwest peak that is 2.5 times brighter than the rest of the azimuthal directions in Band 3, suggesting that dust accumulates not only radially but also azimuthally outside the planets.
  \item The lower limit of the dust mass is 28 $M_{\mathrm{Earth}}$ in the whole ring and 9.3 $M_{\mathrm{Earth}}$ in the northwest peak. This indicates that there is sufficient dust to form additional solid planets or a core of a gas giant.
  \item The central emission exhibits different morphologies: it is extended in Band 7, while it appears as a point source in Band 3. The low spectral index of $\alpha_{\mathrm{SED}} < 2$ suggests that the Band 3 emission is dominated by free-free emission, whereas the Band 7 emission is dominated by thermal dust emission.
  \item The circumplanetary disk found in Band 7 was not detected in Band 3. We put a lower limit of $\alpha_{\mathrm{SED}} > 1.5$.
  \item We estimated the eccentricity of the outer ring and found that the ring is circular. In comparison with numerical simulations, we constrained the mass of PDS 70 c to be less than $ 4.9 M_{\mathrm{Jupiter}}$, assuming the gas turbulence strength $\alpha_{\mathrm{turb}} = 10^{-3}$.
  \item The existence of the planets and the localized dust accumulation suggest that dust accumulates radially and azimuthally outside the planets, promoting inside-out planet formation.
\end{enumerate}

\section*{Acknowledgments}

The authors thank the anonymous referee for the helpful comments and suggestions that significantly improved the manuscript.
This paper makes use of the following ALMA data: ADS/JAO. ALMA\#2015.1.00888.S, 2018.A.00030, 2022.1.00893.S, and 2022.1.01477.S.
ALMA is a partnership of ESO (representing its member states), NSF (USA), and NINS (Japan), together with NRC (Canada), MOST and ASIAA (Taiwan), and KASI (Republic of Korea), in cooperation with the Republic of Chile. 
The Joint ALMA Observatory is operated by ESO, AUI/ NRAO, and NAOJ.
This work was supported by JSPS KAKENHI Grant Numbers JP22K03680, JP22KJ1435, JP23K03463, JP23KJ0636, and JP23KJ1008.
H.B.L. is supported by the National Science and Technology Council (NSTC) of Taiwan (Grant Nos. 111-2112-M-110-022-MY3, 113-2112-M-110-022-MY3).
MB received funding from the European Research Council (ERC) under the European Union's Horizon 2020 research and innovation programme (PROTOPLANETS, grant agreement No. 101002188).

Facility: ALMA.

Software: 
\textit{CASA} \citep{McMullin2007},
\textit{astropy} \citep{astropy2022}, 
\textit{emcee} \citep{Foreman-Mackey2013},
\textit{scipy} \citep{scipy2020},

\appendix

\section{Image Fitting}\label{app:ring_fit}

We perform image fitting using \texttt{ring\_fit} \citep{Doi2023} to examine the morphology of the continuum emission.
Here, we fit the observed emission with two components, the ring and the central emission.
We model the ring with a Gaussian ring given by
\begin{equation} \label{eq:_PDS70_ring}
  I(r) = I_0 \exp\left(-\frac{(r-r_0)^2}{2w^2}\right),
\end{equation}
where $I_0$ is the peak intensity, $r_0$ is the distance of the ring, $w$ is the width of the ring. 
For the central source, we consider a Gaussian disk given by
\begin{equation} \label{eq:_PDS70_point}
  I(r) = I_0 \exp\left(-\frac{r^2}{2w^2}\right),
\end{equation}
where $I_0$ is the peak intensity, $w$ is the width of the emission. 
In other words, equation (\ref{eq:_PDS70_point}) is the same as equation (\ref{eq:_PDS70_ring}) with $r_0 = 0$.
The caveat is that the emission of the object is not perfectly Gaussian, but the aim here is not to reproduce the emission shape perfectly but to quantitatively evaluate the position and the spatial distribution of the emission.

\begin{figure*}[phtb]
  \centering
  \includegraphics[width=0.9\textwidth]{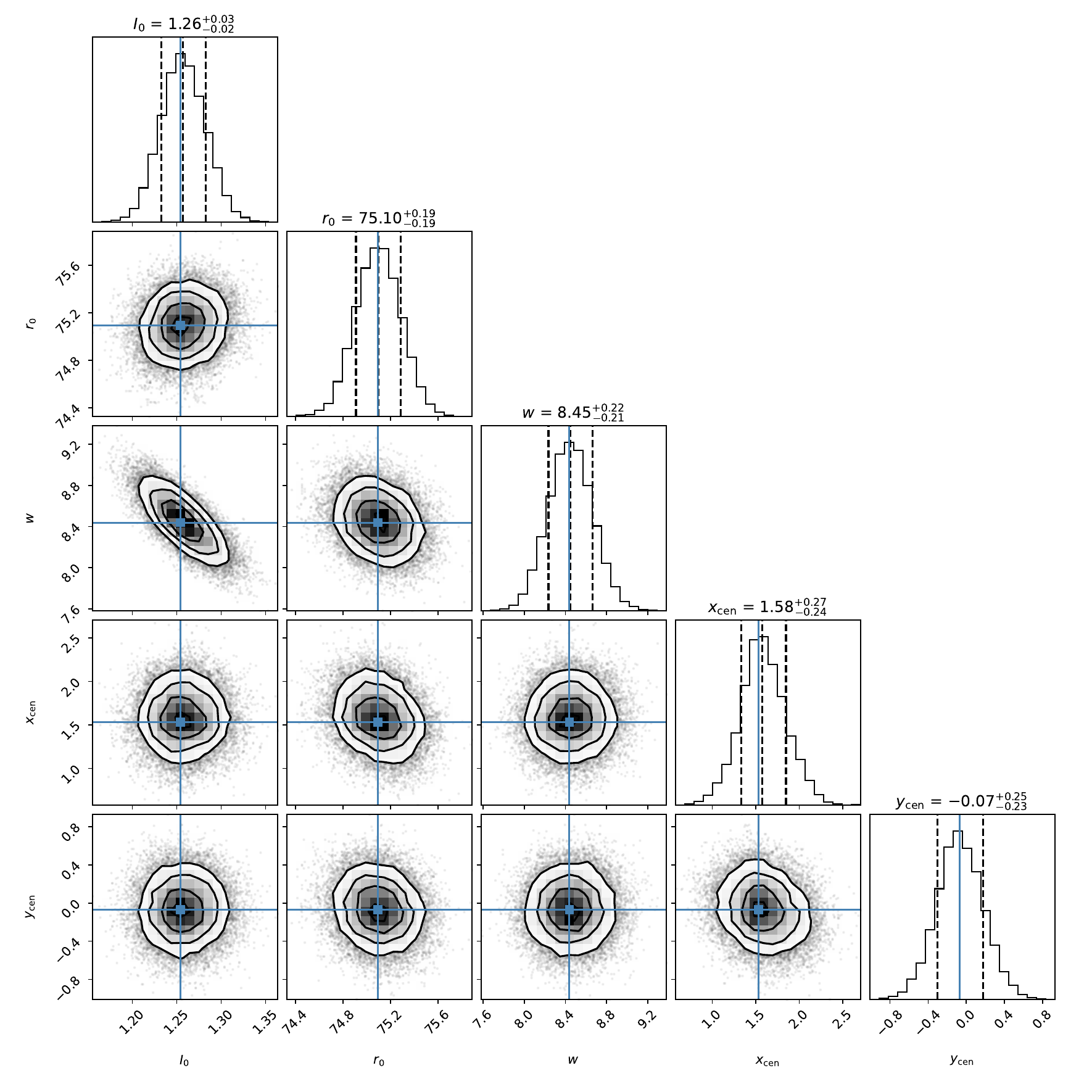}
    \caption{
    The corner plot of MCMC fitting for the ring of PDS 70 in Band 3. The dashed lines in the top panels show the 16th, 50th, and 84th percentiles of the marginal distributions. The blue lines show the parameters that maximize the joint probability.}
  \label{fig:MCMC_B3_ring}
\end{figure*}

\begin{figure*}[phtb]
  \centering
  \includegraphics[width=0.9\textwidth]{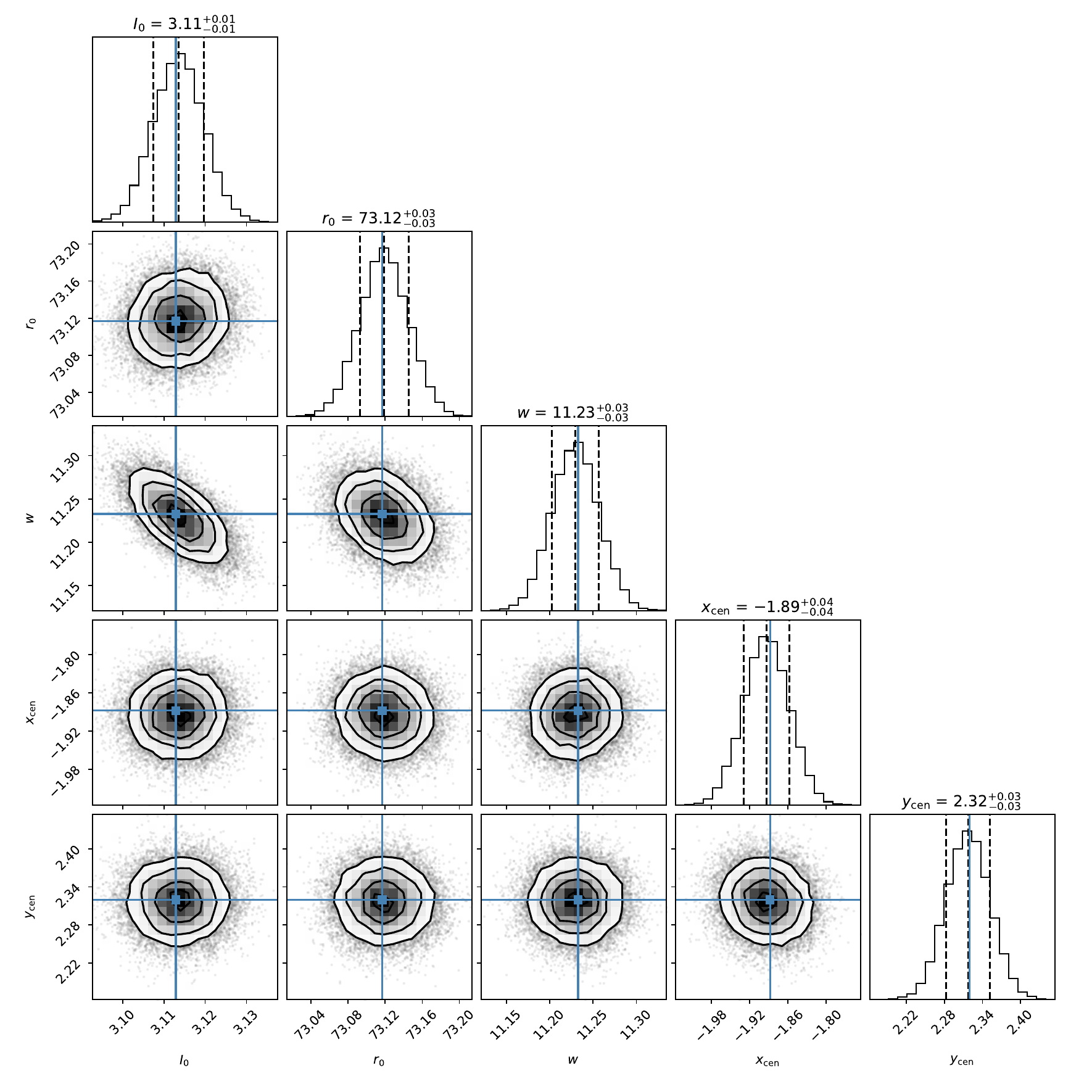}
    \caption{
      The same as Figure \ref{fig:MCMC_B3_ring}, but in Band 7.
    }
    \label{fig:MCMC_B7_ring}
\end{figure*}

\begin{figure*}[phtb]
  \centering
  \includegraphics[width=0.75\textwidth]{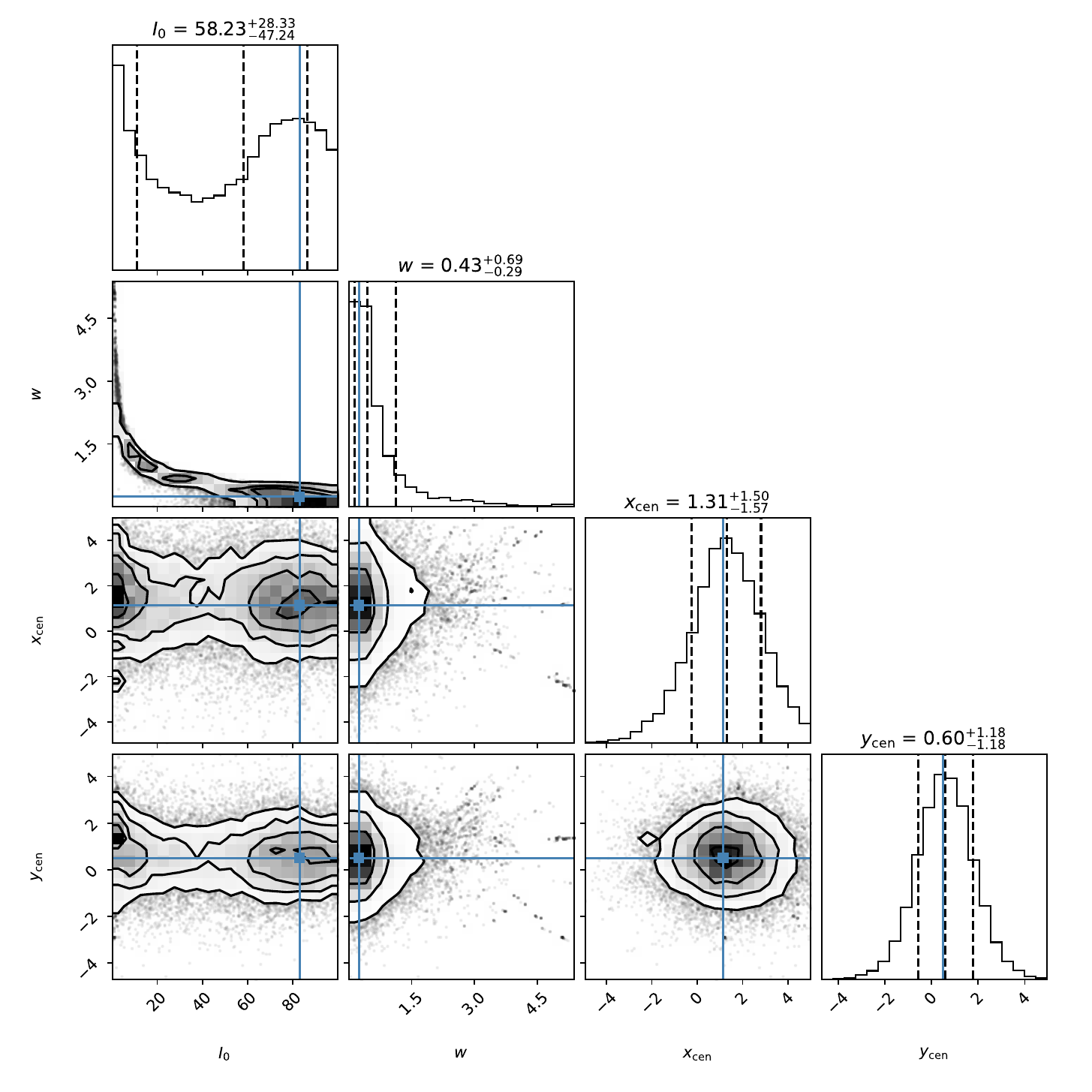}
    \caption{
    The same as Figure \ref{fig:MCMC_B3_ring}, but for the central source in Band 3.
    Since the emission is smaller than the beam size, the peak intensity and size are degenerate.
    }
  \label{fig:MCMC_B3_point}
\end{figure*}

\begin{figure*}[phtb]
  \centering
  \includegraphics[width=0.75\textwidth]{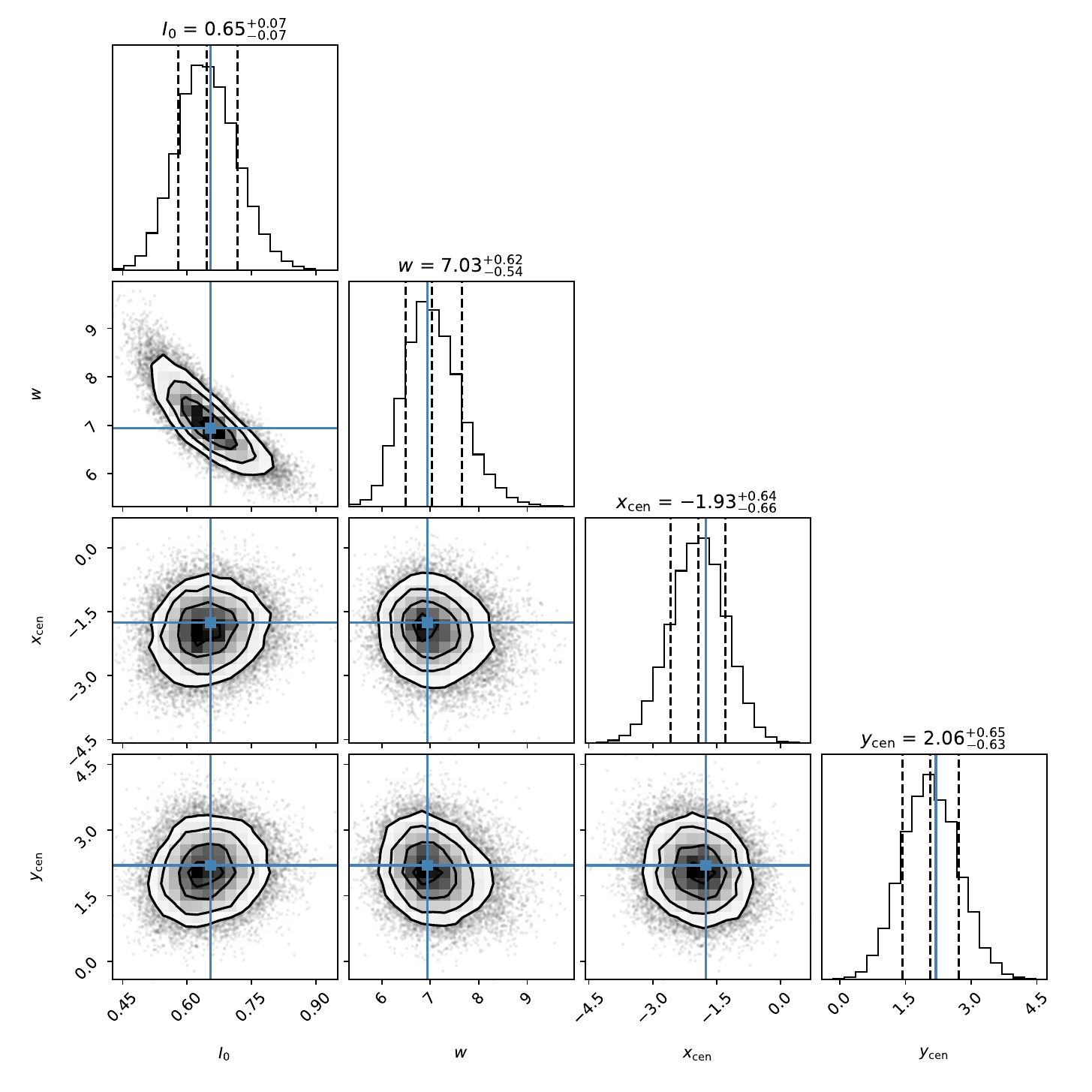}
    \caption{
    The same as Figure \ref{fig:MCMC_B3_ring}, but for the central source in Band 7.
    Since the emission is spatially resolved, the peak intensity and size are not degenerate as in Band 3.
    }
    \label{fig:MCMC_B7_point}
\end{figure*}

We created these model images, smoothed them with the beam, and minimized the residuals between models and observations using \texttt{emcee} \citep{Foreman-Mackey2013}.
The free parameters are the parameters of equations (\ref{eq:_PDS70_ring}) or (\ref{eq:_PDS70_point}) and the center positions of these structures in the deprojected plane $(x_{\mathrm{cen}}, y_{\mathrm{cen}})$.
We fixed the inclination $i = 51.7^{\circ}$ and the position angle PA = $160.4^{\circ}$ based on \citet{Keppler2019}.
We fitted the ring in the range of $35\ \mathrm{au} < r < 105\ \mathrm{au}$ and the central source in the range of $r < 15$ au.
We performed fitting for the full azimuthal angle and also performed the fitting for the range of $0^{\circ} < \phi < 270^{\circ}$ excluding the northwest peak, since the non-axisymmetric nature of the northwest peak could affect the fitting.
We perform MCMC fitting with 16 walkers and 20,000 steps, excluding the first 1,000 steps as burn-in.
Figure \ref{fig:MCMC_B3_ring} and \ref{fig:MCMC_B7_ring} show the corner plots of the fitting results for the ring in Band 3 and Band 7, respectively, and Figure \ref{fig:MCMC_B3_point} and \ref{fig:MCMC_B7_point} show those for the central source in Band 3 and Band 7, respectively.
The fitting results are summarized in Table \ref{tab:PDS70_fitted}.
The results excluding the northeast peak show a slight decrease in the peak intensity, but no significant difference in the center position.
In the main discussion, we use the fitting results for full azimuthal angle, while the conclusions do not change depending on which results are used.

\begin{figure}[htbp]
  \centering
  \includegraphics[width=0.4\textwidth]{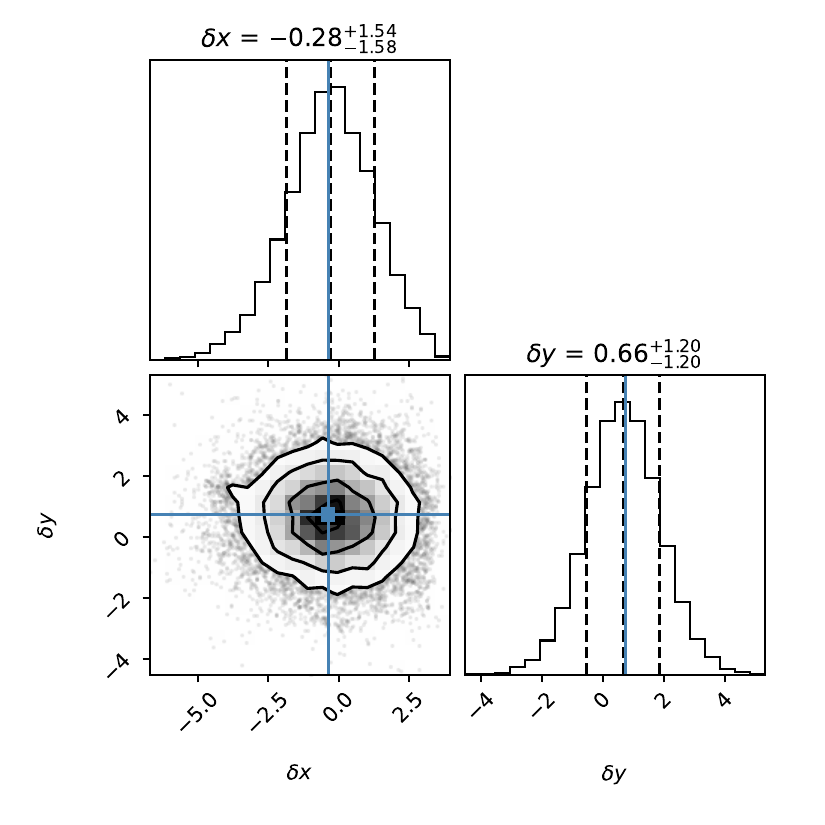}
    \caption{
    The corner plot of the difference in the center position between the ring and the central source in Band 3.
    It shows that the center of the ring and the central source are consistent.
    }
  \label{fig:MCMC_B3_center_shift}
\end{figure}

\begin{figure}[htbp]
  \centering
  \includegraphics[width=0.4\textwidth]{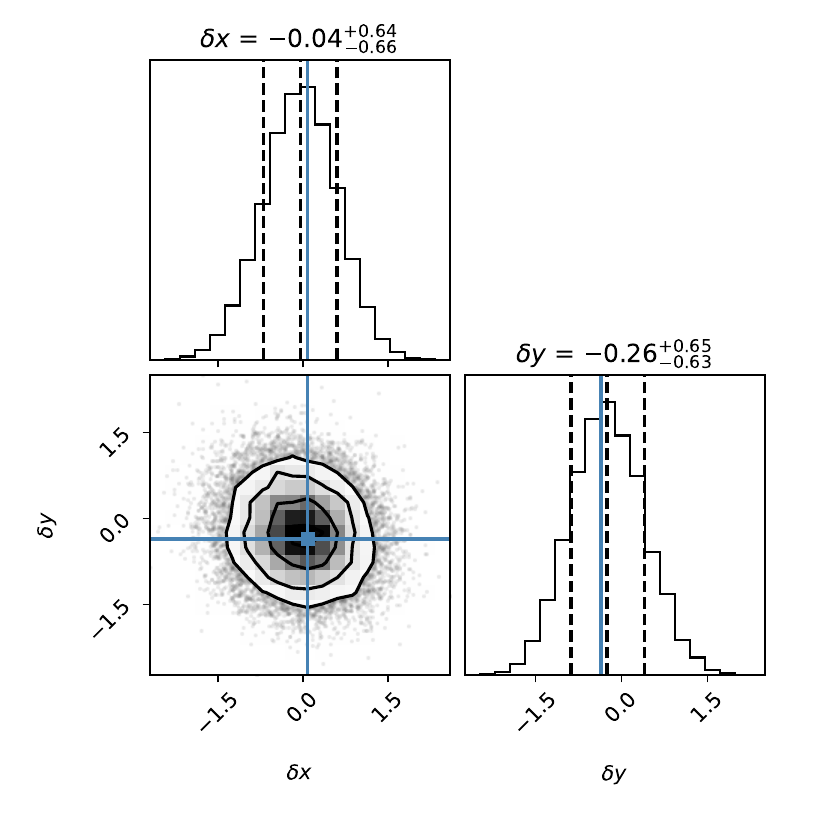}
    \caption{
      The same as Figure \ref{fig:MCMC_B3_center_shift}, but in Band 7.
    }
    \label{fig:MCMC_B7_center_shift}
\end{figure}

To determine the eccentricity of the ring, we calculated the difference of the center positions between the ring and the central source from the fitting results.
We calculate the offset of the center positions as 
\begin{equation}
  \delta \{x, y\} = \{x, y\}_{\mathrm{cen,central}} - \{x, y\}_{\mathrm{cen,ring}},
\end{equation}
and show the corner plots of these differences in Figure \ref{fig:MCMC_B3_center_shift} and Figure \ref{fig:MCMC_B7_center_shift}.
These results show that the center of the ring and the central source is consistent, i.e., the ring is circular.
Furthermore, we calculate the offset of the center positions between the ring and the central source as $\delta r = \sqrt{\delta x^2 + \delta y^2}$ and set the 84th percentile as the upper limit of the deviation of the ring center position.
These results are summarized in Table \ref{tab:PDS70_fitting_center}.

\begin{deluxetable*}{ccccccc}
  \tablecaption{Fitted parameters for each component\label{tab:PDS70_fitted}}
  \tablehead{
    \colhead{Band} & \colhead{Component} & \colhead{$I_0$} & \colhead{$r_0$} & \colhead{$w$} & \colhead{$x_{\mathrm{cen}}$} & \colhead{$y_{\mathrm{cen}}$} \\
    & & [mm] & [au] & [au] & [au] & [au]
  }
  \startdata
  \multirow{3}{*}{B3} 
  & ring   & $1.26_{-0.02}^{+0.03}$ & $75.10_{-0.19}^{+0.19}$ & $8.45_{-0.21}^{+0.22}$ & $1.58_{-0.24}^{+0.27}$ & $-0.07_{-0.23}^{+0.25}$ \\
  & $\mathrm{ring_{exclude}}$ & $1.03_{-0.03}^{+0.03}$ & $74.68_{-0.30}^{+0.30}$ & $9.50_{-0.30}^{+0.29}$ & $0.77_{-0.42}^{+0.41}$ & $-0.25_{-0.38}^{+0.39}$ \\
  & center & $(58.23_{-47.24}^{+28.33})$ & - & $<1.12$ & $1.31_{-1.57}^{+1.50}$ & $0.60_{-1.18}^{+1.18}$ \\
  \hline
  \multirow{3}{*}{B7}
  & ring  & $3.11_{-0.01}^{+0.01}$ & $73.12_{-0.03}^{+0.03}$ & $11.23_{-0.03}^{+0.03}$ & $-1.89_{-0.04}^{+0.04}$ & $2.32_{-0.03}^{+0.03}$ \\
  & $\mathrm{ring_{exclude}}$ & $2.98_{-0.01}^{+0.01}$ & $73.45_{-0.04}^{+0.04}$ & $11.42_{-0.03}^{+0.03}$ & $-1.60_{-0.05}^{+0.05}
$ & $2.80_{-0.05}^{+0.05}$ \\
  & center & $0.65_{-0.07}^{+0.07}$ & - & $7.03_{-0.54}^{+0.62}$ & $-1.93_{-0.66}^{+0.64}$ & $2.06_{-0.63}^{+0.65}$ \\
  \enddata
  \tablecomments{The results of the MCMC fitting. Since the width and intensity of the central point source in Band 3 are degenerate, we show the upper limit of the width, and the intensity has no useful constraints. $\mathrm{ring_{exclude}}$ is the fitting result of the ring excluding the range of $270-360^{\circ}$ where the northwest peak is located.}
\end{deluxetable*}

\begin{deluxetable}{cccc}
  \tablecaption{Offset of the center\label{tab:PDS70_fitting_center}}
  \tablehead{
    \colhead{Band} & \colhead{$\delta x$} & \colhead{$\delta y$} & \colhead{$\delta r$} \\
    & [au] & [au] & [au]
  }
  \startdata
  B3 & $-0.28_{-1.58}^{+1.54}$ & $ 0.66_{-1.20}^{+1.20}$ &  $< 2.88$ \\
  B7 & $-0.04_{-0.66}^{+0.64}$ & $-0.26_{-0.63}^{+0.65}$ & $< 1.28$ \\
  \enddata
  \tablecomments{The offset of the center positions between the ring and the central source. This difference corresponds to the offset of the focus from the center, i.e., the eccentricity.}
\end{deluxetable}

\section{Radial Profile}\label{app:radial_profile}

\begin{figure*}[htbp]
  \centering
  \includegraphics[width=0.75\textwidth]{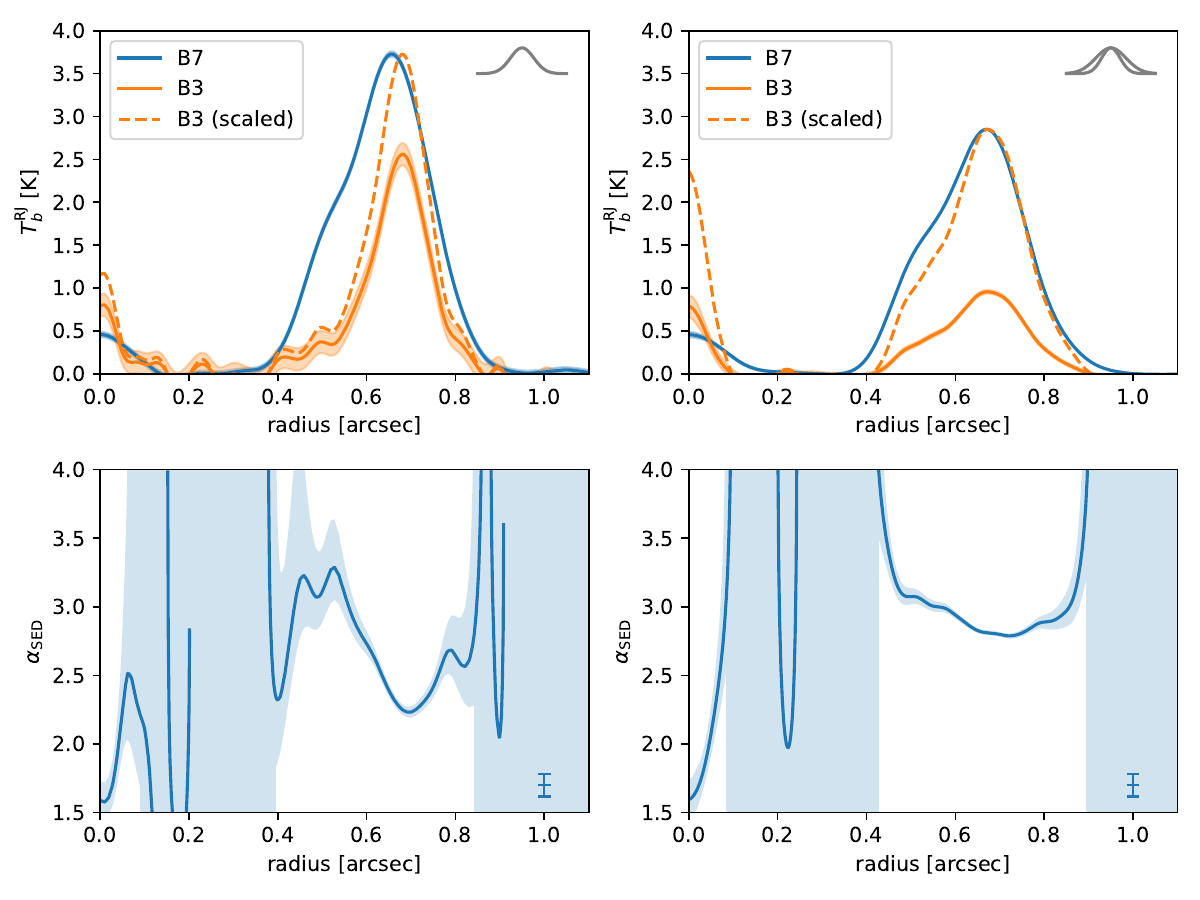}
    \caption{
      The same as Figure \ref{fig:PDS70_radial}, but for the radial profiles in the northwest peak direction (left) and the azimuthally averaged radial profiles in the directions other than the peak (right). The upper panels show the radial intensity profiles, while the lower panels show the spectral index profiles. The Gaussian in the upper right corner of the upper left panel represents the beam size in the northwest peak direction, and the Gaussian in the upper right corner of the upper right panel shows the major and minor axes of the deprojected beam.
    }
    \label{fig:radial_4panel}
\end{figure*}

The radial profile shown in Figure \ref{fig:PDS70_radial} is the azimuthally averaged profile that includes emission from both the northwest peak and the other directions.
Here, we show the radial profiles for the northwest peak direction and the rest directions separately.
The left panel of Figure \ref{fig:radial_4panel} shows the radial profile in the northwest peak direction, PA = 309$^{\circ}$, and the right panel shows the azimuthally averaged profile in the directions other than the northwest peak, 0$^{\circ} < \mathrm{PA} < 270^{\circ}$.
The upper panels show the radial profiles, and the lower panels show the spectral index profiles as in Figure \ref{fig:PDS70_radial}.

The profile in the northwest peak direction has larger noise because it is not azimuthally averaged, but the peak intensity is higher in Band 3 than in the other directions.
The intensity of the shoulder relative to the radial peak is smaller in Band 3 than in Band 7, but the noise in the shoulder region makes it difficult to quantitatively discuss the contrast between the peak and the shoulder.
The spectral index shows a minimum at the center of the ring.

The radial profile in the directions other than the northwest peak also shows the ring and the inner shoulder, and the intensity of the shoulder relative to the radial peak is smaller in Band 3 than in Band 7.
Thus, the discussion based on the radial profile in Section \ref{sec:PDS70_ring} is qualitatively valid for all azimuthal directions.

\bibliography{doi_citation_phdthesis}{}
\bibliographystyle{aasjournal}

\end{CJK*}
\end{document}